# PREVIS - A Combined Machine Learning and Visual Interpolation Approach for Interactive Reverse Engineering in Assembly Quality Control

Patrick Ruediger[a,*], Felix Claus[b], Viktor Leonhardt[c], Hans Hagen[b], Jan C. Aurich[a], Christoph Garth[c]

[a]*Institute for Manufacturing Technolgy and Production Systems (FBK), Technische Universität Kaiserslautern, P. O. Box 3049, 67653 Kaiserslautern, Germany*
[b]*Computer Graphics & HCI Lab, Technische Universität Kaiserslautern, P. O. Box 3049, 67653 Kaiserslautern, Germany*
[c]*Scientific Visualization Lab, Technische Universität Kaiserslautern, P. O. Box 3049, 67653 Kaiserslautern, Germany*

* Corresponding author. Tel.: +49-631-205-4282; fax: +49-631-205-3304. E-mail address: patrick.ruediger@mv.uni-kl.de

**Abstract**

We present PREVIS, a visual analytics tool, enhancing machine learning performance analysis in engineering applications. The presented toolchain allows for a direct comparison of regression models. In addition, we provide a methodology to visualize the impact of regression errors on the underlying field of interest in the original domain, the part geometry, via exploiting standard interpolation methods. Further, we allow a real-time preview of user-driven parameter changes in the displacement field via visual interpolation. This allows for fast and accountable online change management. We demonstrate the effectiveness with an ex-ante optimization of an automotive engine hood.

## 1. Introduction

The stunning results and fast development of deep learning techniques enabled machine learning transfer in various application scenarios in the last decade. Most engineering applications models are mainly designed explicitly for a narrow application field (e.g., visual object detection in a production line, for specific single our small group of predefined object features). Consequently, such a model's development is very time-consuming and only profitable when daily decisions justify the initial and ongoing development costs, even under changing production parameters. Today's production systems have an increased variety in products [1] which makes those highly specialized models unprofitable in many applications, and a human workforce is preferred.

However, tools for quickly building machine learning systems have generally outpaced the growth and adoption of tools to understand whether they are reliable.

For instance, low average error on initial training data does not give assurances under data set shift [2]. A model with low average error can also still make significant pointwise errors on individual predictions for some types of inputs [3].

These shortcomings are the motivation for the development of the presented tool PREVIS. It is designed explicitly for regression problems, which map a field on a geometry to a set of process parameters. Standard visualization and analysis techniques for these problems solely analyze the regression error on the predicted parameters. The linking to the geometry's domain space happens only through an interpretation of the sensitivity analysis results. In contrast, we strive to directly analyze regression performance on both the predicted parameters and the resulting field with the proposed approach. Our key contributions can be summarized as follows:

- Comparing prediction performance of multiple regression models at once.
- Mapping the prediction error from parameter space to domain space (on the part geometry, see Figure 1).
- Allowing specific test queries on the parameters to evaluate their impact on the resulting fields separately.

We demonstrate the usage of our tool with an example from the ex-ante assembly quality control in the automotive industry.

*1.1. Related Work*

From the perspective of the end-user, machine learning is still mostly labeled as a black box. Throughout the years, visual analytics is approaching this topic, intending to color this black box. In the work of Duch [4], he made some early attempts that highlight the four dominant effects in learning models, such as neural networks. These are the dynamics of neural learning, under and over-fitting effects, regularization effects, and differences between networks with the same performance. Tzeng et al. [5] extended this work intending to improve such networks' engineering. With the increased usage of deep neural networks (DNNs) in the last years, the black box coloring on the level of single neurons is no longer a feasible approach due to the immense number of neurons used here.

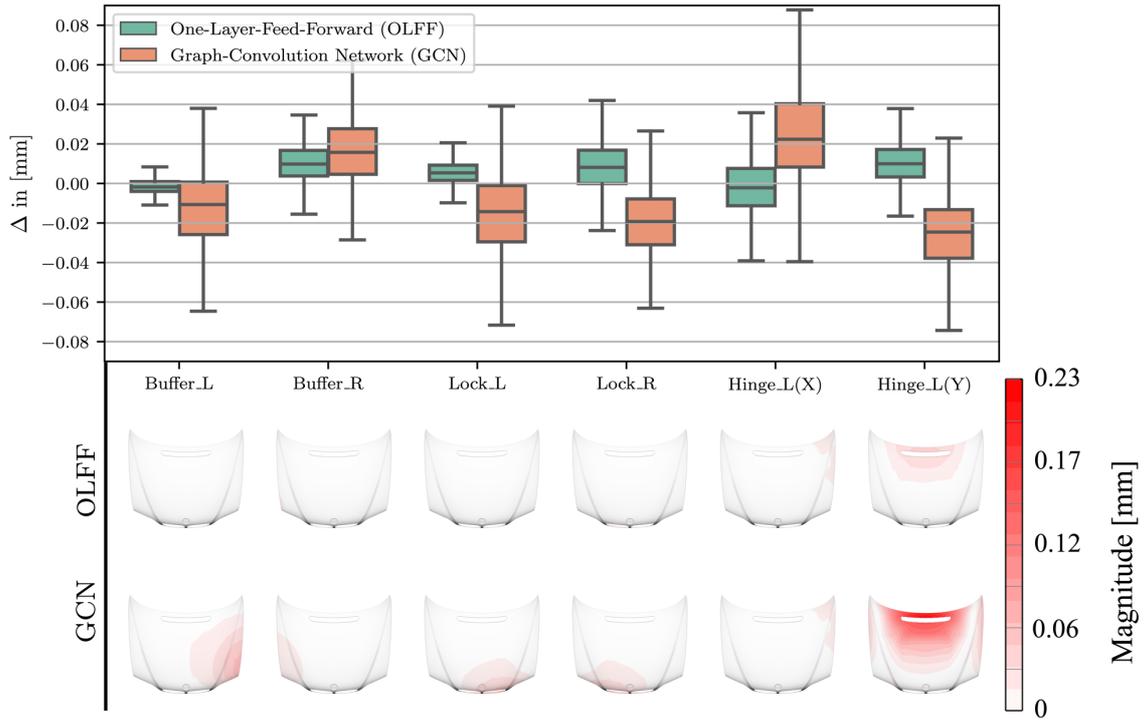

Figure 1. Comparing ML-Model Performance in both ways. In the top view we directly evaluate the prediction error for each parameter and model using a boxplot. In the bottom view, we used the direct interpolation via LLSF to map the predictive errors back on to the displacement field on the part geometry. Here we can analyze the differences of the prediction models in more detail and spot differences and commonalities. The errors shown are always relative errors.

As a consequence, the visual analytics approach switched more to the output side. Samek et al. [6] for example, introduced a direct visualization approach, which highlighted the learned features of images using the three techniques of a sensitivity analysis [7], Deconvolution Method [8], and the LRP Algorithm [9]. Inspired by the focus on the output performance, Smilkov et al. [10] introduced a direct-manipulation visualization approach, which communicates the method of neural networks via an interactive playground rather than coding. Focusing on the output, Zintgraf et al. [11] present the prediction difference analysis method for visualizing the response of a neural network. While this work is focused mainly on image classification, it contains the base idea of our visual analytics pipeline for the regression task. Finally, for the combination and comparison of multiple models and input data, Chatzimparmpas et al. [12] introduced StackGenVis.

## 2. Use-Case

The selected showcase is the process of assembling an engine hood (Figure 2) to its chassis. For this use-case scenario, two different machine learning (ML) approaches were developed to estimate the engine hood's boundary states based on a post-assembly 3D-Scan of its visible skin. PREVIS is used to qualify and compare the different approaches beyond conventional performance indicators like ROC curves or percentage precision values.

### 2.1. Simulation Setup

The training data used by the ML approaches was obtained by generating FE-simulation ensembles. The used car hood itself is an assembly containing seven individual sheet metal parts, connected by spot welds and different types of adhesives (see Figure 2).

Based on the CAD files, the simulation model was created by meshing the geometry with first-order 3D-shell elements

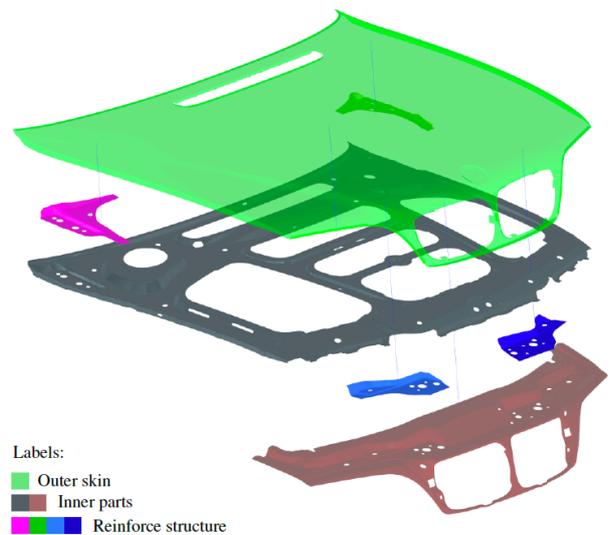

Figure 2. Overview of the whole assembly structure of the engine hood, which is used in this article. The colors of parts in the exploded view drawing separate the individual metal sheets.

(total number of elements: 365.154) and connecting the assembly considering spot welds and adhesive positions and behavior as well as material and thicknesses of individual components. The material model (steel) is assumed to be linear-elastic with an e-modulus of 210 Gpa and a Poisson rate of 0.3. Two fixed, external loads, modeling the gas springs near the hinges with the magnitude of 580 N each, complete the model. The presented hood has two hinges, two locks, and two buffers as mechanical boundaries attaching the hood to the chassis, see Figure 3. The engine hood's boundaries are adjustable to deal with production uncertainties and tolerances during production. For the present use case, six boundaries are considered, which are adjustable and vary within the interval of [-1mm,1mm] in the simulation model. The simulation output of each run is a 3D vector displacement field defined on the vertices of the mesh. Two different simulation ensembles are created. Both ensembles sample the parameter space, spanned by the six adjustable parameters evenly using Latin hypercube sampling with the sample size of 1400 for the test set and at the corners and edges of the parameter space for the training 729 entries in the training set.

## 2.2. Interpolation with PCA and LLSF

For mapping an arbitrary set of parameters to the solution space, an LLSF with a basis generated by PCA performed on the training data is chosen. The generation of the basis via PCA will be briefly described in the following.

A PCA is performed on the training data set's solution space to obtain an optimal basis for the LLSF. As the PCA cannot handle vector fields, the simulation results are compressed by condensing the displacement vector fields to scalar fields. It is achieved by calculating the displacement magnitude per node in the surface normal direction (signed magnitude). The PCA can then decompose the resulting 729 scalar fields into seven principal components (PCs). Adding more PCs does not change the amount of information carried by the basis in a relevant way as most information is carried by PCs 1-7. To ensure not lose the information we, however, decided to add three more PCs. Together the first 10 PCs explain 99.99985% of the variance in the current data set.

The outputs of the PCA are weights for each of the 729 input scalar fields per PC. A linear combination on the parameter space is calculated for each PC with these weights, resulting in one displacement field per PC. The same linear combination is also calculated in the parameter space. Thus, ten displacement fields with ten corresponding parameter sets are obtained that can be used to explain the whole simulation ensemble. Note that the generated basis displacement fields are artificially created and do not represent any performed simulation run. With this PCA-generated basis, an LLSF can be performed using any parameter set as a target resulting in an estimation of the corresponding displacement field. Note that this method cannot used for interpolation in general and works only in special cases where the number of needed PCs is small enough that the LLSF performed on the parameter space converges to the desired result. For the shown application, the approximation of a displacement field based on a given parameter set was tested with the whole validation data set and works satisfyingly well, showing maximum errors of below 0.1 mm at single mesh nodes. For the general case, we would advise using multivariate interpolation methods performed on the simulation ensembles. If even this is not possible due to high estimation errors, consistently adequate simulation runs can be performed with the cost of much higher run times.

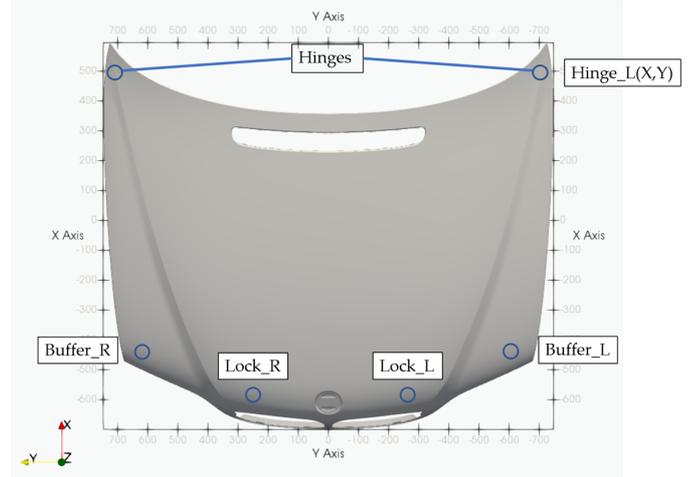

Figure 3. Sheet metal car body part used in automotive industry. The used boundaries for the simulation are highlighted, with external forces used via gas springs. The coordinate system for this part is shown.

## 2.3. ML-based predictive models

The introduced ML-models are solving a regression task in which they predict a set of simulation parameters based on an input displacement field

### 2.3.1. One-Layer-Feed-Forward Network

Yanez-Marquez et al. [13] proposed the minimalist machine learning approach on the quest for bleaching the black box in machine learning. As a consequence, a fully connected one (hidden) layer model is developed. These networks can learn polynomial functions of degree r over d-dimensional inputs [14] and are therefore well-suited for the linear-elastic deformations in the shown use-case. Based on a parameter study, we found out that using 75 neurons for the hidden layer is a good trade-off between over-parameterization and accuracy. With 75 neurons in a fully connected network, we end up with 42,426,306 trainable parameters for the shown use case. In order to cope with the effect of over-parameterization, we use the stochastic gradient descent (SGD) with the Nesterov Momentum [15] as this optimizer is the most robust in this kind of setting, as shown by Brutzkus et al. [16]. After 2000 epochs, the trained network is entering a stable state, which we will use as the reference model for comparison here.

### 2.3.2. Graph Convolution Neural Network

The success of convolutional neural networks has driven the generalization of this architecture to more general domains. A graph convolution neural network, like described by Defferrard et al. [17], operates on a graph $G = (V, A)$ and a signal $s$ on its vertices $V$. The convolution itself utilizes the Fourier basis $U$, which can be computed by the eigenvalue decomposition $L = U \Lambda U^T$. At the same time $L$ represents the normalized Laplacian matrix of the graph $G$. A convolution on the CAD files can be achieved by interpreting the meshed geometry as an undirected graph, including the 3D displacement vector of each vertex. Thus, the neural network operates on the whole displacement field and includes the relations of all vertices.

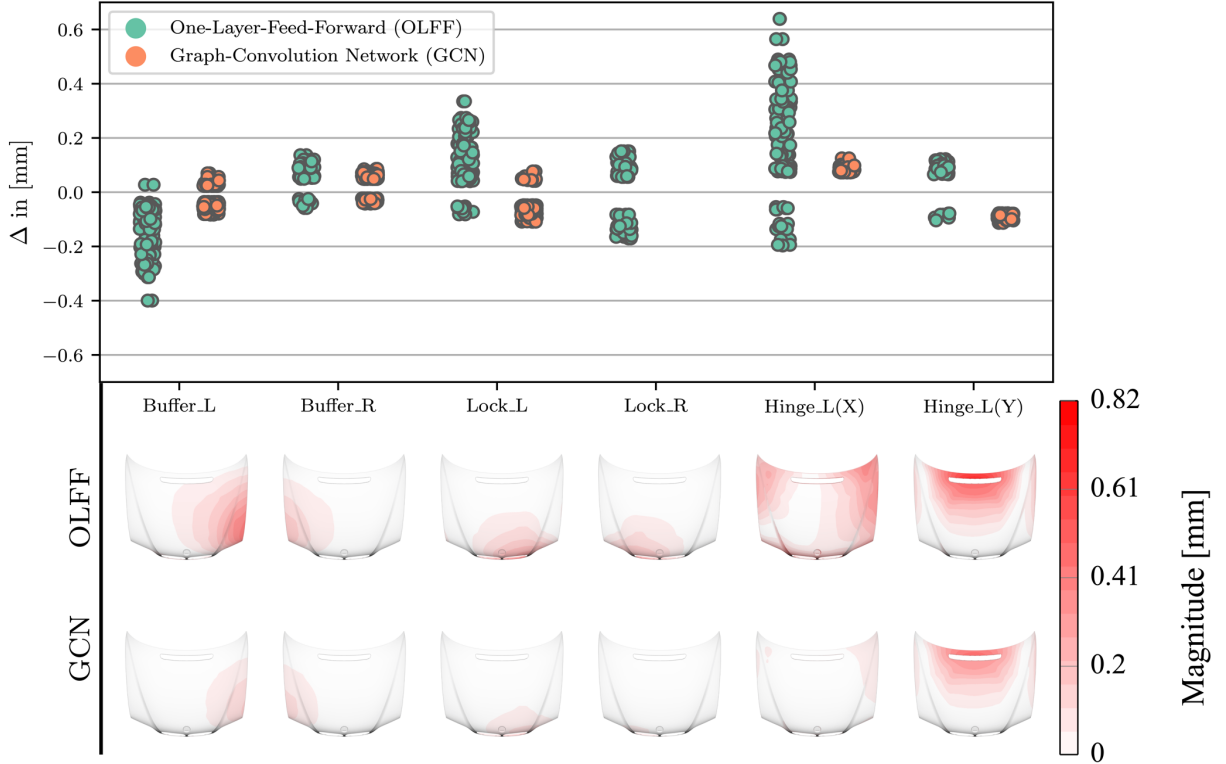

Figure 4 Outlier Analysis. We can use the introduced direct interpolation method to analyze the outliers of the prediction models in more detail. In the top view we can observe that the GCN model is more stable than the OLFF model, as we spot fewer and less cluttered outliers. In the interpolated bottom view, we can now analyze the impact of these outliers on the displacement field on the part geometry. Here we can observe similar patterns for both models, but the GCN is better handling them.

The used architecture consists of three layers: (1) graph convolution, (2) fully connected layer and (3) output layer. The graph convolution is slightly modified to operate on the first $\mu$ smallest eigenvectors (Fourier basis) and neglects the inverse Fourier transformation. Besides, the Fourier basis is computed explicitly, instead of approximating it using the Chebyshev polynomials, and the Chebyshev polynomials are modified to enforce a convergence to zero at the $\mu$th frequency. Thus, the followed fully connected layers approximate the six boundaries using a continuous function [14] in the spectral domain. The core concept of this architecture is to build a space in the spectral domain using the trained filters of the graph convolution, which represents the whole data, and to solve the approximation within this space as a regression.

The hyperparameters *user-driven parameters* used in this work are as followed:
- The graph convolution operates on 25 filters of the first 100 frequencies, while 15 Chebyshev polynomials of the first kind are used to describe the filters in their spectral representation.
- The number of neurons of the first fully connected layer is 2048, and the non-linearity rectified linear unit is applied.
- The last fully-connected layer equals the number of boundaries and uses a bias as an additional trainable variable.

The optimizer AdaGrad [18] is used to find the solution to this regression problem in 300 epochs.

## 3. PREVIS Framework

The main idea of the developed framework is to extend the interpretability of regression models $\mathcal{P}$ that map a field $\mathcal{U} \subseteq \mathbb{R}^m$ defined on a manifold $\mathcal{M} \subseteq \mathbb{R}^3$ to a quantity of attributes $\mathcal{A}$.

$$\mathcal{P}: \mathcal{U} \to \mathcal{A}, \quad \mathcal{A} \subseteq \mathbb{R}^n \tag{1}$$

A finite element (FE) simulation is a typical example for engineering applications. Here the manifold is a discretization of the part geometry. The resulting fields (e.g., stress or displacement) are the simulation results $\mathcal{U}$, and the to-be-predicted attributes $\mathcal{A}$ are either explicit or implicit parameters of the simulation, indicated by a bar above the name e.g., $\bar{a}$. The model's performance is then evaluated based on the predicted attributes' correctness (see Figure 1). Interpreting these results can be a challenging task on certain occasions, e.g.:
- For a high number of attributes
- For attributes with higher dimensions (vectors)
- For strongly correlated attributes
- For attributes with varying impact on the resulting field
- For comparing models with similar behavior

In general, it is hard to find implications for model improvement or the model's generalization capabilities based on this narrow analysis. As our initial problem set is defined on a manifold, we would like to incorporate it in our model performance. Therefore, we need an interpolation method that maps the impact of the prediction performance on the manifold. For this, we are using the Linear-Least-Squares-Fit (LLSF) to map a set of attributes back on the manifold. The LLSF is an interpolation method $\mathcal{I}$, which yields an error $\varepsilon$. Let $U \in \mathcal{U}$ be a field which attributes $\bar{a} \in \mathcal{A}$ are known and an interpolated field $U^*$, which results by applying the interpolation to a set of attributes $\bar{a}$.

$$\mathcal{I}: \mathcal{A} \to \mathcal{U} \tag{2}$$

$$U^*(\bar{a}) = U(\bar{a}) + \varepsilon(\bar{a}) \tag{3}$$

Using Equation (3) we now can calculate the difference between an interpolated prediction $U^*(a)$ and test set $U^*(\bar{a})$.

$$\Delta U = U^*(a) - U^*(\bar{a}) \tag{4}$$

$$\Delta U = U(a) - U(\bar{a}) + (\varepsilon(a) - \varepsilon(\bar{a})) \tag{5}$$

This serves now as a relative measure for the model's performance as the interpolation error is vanishing under certain conditions.

For each difference in the predicted parameters, which is analyzed using the box plots from Figure 1, we now interpret $\Delta U$ as the induced relative change on the resulting field. Thus, we can map the models' prediction performance back on the manifold, which allows for more profound and detailed analysis. But this mapping is only applicable when:
- The overall interpolation error of the chosen method $\varepsilon$ is vanishing: $\varepsilon(a) - \varepsilon(\bar{a}) \to 0$
- The variance of the interpolation error is small, such that $\varepsilon(a) - \varepsilon(\bar{a}) \to c_\varepsilon$ can be assumed constant.

As long we can define an interpolation method that fulfills these requirements, our methodology is applicable for any application in the described problem set. In the shown use cases, we use the LLSF on a precalculated Principal Component Analysis (PCA) - Basis.

## 4. Results

At first, we compare the different models against each other. In regression problems, we typically use the difference to the actual target value as our measurement. Given the six parameters of our use case, we can then use a boxplot to summarize our models' predictive errors for all parameters and compare them (see Figure 1, top). We can then extend this view by applying our method to view the same key statistic features (mean, variation, min, max) on the part geometry. In this particular case, each approach's whiskers' span is visualized on the 3D-domain (see Figure 1, bottom). Note that in this Figure, the outliers of the boxplot are excluded. The mapping of estimation errors back to the 3D domain allows us to evaluate each parameter's prediction error's impact on the resulting displacement. Using this visualization, the differences between the two ML models got clearer. The One-Layer-Feed-Forward (OLFF) model does, in general, perform better compared to the Graph-Convolution-Network (GCN). This statement can be underlined by both - the boxplot whiskers and the corresponding impacts to the resulting displacement field. However, in the visualizations of the error impact, it can be observed that especially the parameter Hinge_Y is highly sensitive. At the same time, Hinge_X does show almost the same impact for totally different deviation seen in the boxplot.

Next, we take a deeper look into Figure 4. Here, only the outliers of the estimation errors are shown. Note that the error range in parameter space increased by a factor of 10, while for the induced error on the displacement field, it increased by a factor of 3 to 4, which results in a non-linear dependency of the two error spans. The corresponding impact plots show the impact of the entire error span per parameter. Here we can see that the large error values outside the whiskers' span lead, especially for the OLFF approach to high impacts on the 3D domain. In contrast, the GCN approach only performs slightly worse when including outliers (compared to Figure 1).

In summary, the OLFF approach does perform better for most of the validation data, while the GCN approach provides a more stable estimation across the whole solution space.

Table 1 shows the execution run times for the shown use-case. Critical for the interactive aspect of our method is the execution time of the interpolation. As one interpolation only needs 0.05 s and all interpolation can be performed in parallel, requirements for real-time updates are achieved. By that, for instance, a boxplot can be interactively explored on the 3D domain.

Table 1. Computational run times.

| Task | Time per Execution | # of Execution | Total Time |
| --- | --- | --- | --- |
| Parameter Estimation (OLFF) | 8.314 ms | 1400 | 11.6 s |
| Parameter Estimation (GCN) | 9.256 ms | 1400 | 12.9 s |
| Interpolation with PCA & LLSF | 0.05 s | 12 | 0.6 s |

## 5. Discussion

The introduced methodology extends the state-of-the-art approaches for analyzing the predictive performance of ML-regression models. The current approaches that are going beyond this scope were mainly designed for problem sets revolving around images. While this already covers a wide range of applications, our approach extends the aforementioned to fields of up to n dimensions defined on arbitrary manifolds. Visualizing a field with more than $n = 3$ dimensions is still a challenge, and so this remains a bottleneck for our method too. Further, there is a need for a suitable interpolation or, more generally, an inverse function as described in Section 3. If this restriction is unfulfilled, our analysis's expressiveness on the part geometry is reduced drastically, as the interpolation error moderates the results. The results from the shown methodology are only valid compared with others and cannot be interpreted independently. Consequently, the resulting visualizations are not to be used independently to judge the regression models' performance. However, in combination with an error analysis in parameter space as shown in Figures 3 and 4, an accurate performance analysis is achieved.

## 6. Conclusion

We introduced PREVIS as a concept for combining visual analytics and predictive machine learning with visual interpolation. It addresses the challenges of building reliable and explainable machine learning models, especially under data set shifts. We introduced how a direct interpolation method can increase the explainability of predictive models, as it allows for an inverse mapping of the predicted attributes in the original domain space. Based on an assembly use-case of an automotive engine-hood, we demonstrated the workflow and usage of this technique to compare multiple regression models at once. Moreover, this applies to the predicted attributes and the domain space, e.g., the part geometry. In a single view, we can compare the predictive error in parameter and domain space and distinguish the impact of different machine learning model approaches in more detail.

Additionally, this view allows for a more explainable interpretation of the prediction performance, as we can directly see the effects of the predictive error on the underlying displacement field. In our future work, we strive to deploy a generalized toolkit for applying the introduced method on various problem sets, which are described as a field on a domain and a set of attributes related to those fields. With this general approach, we aim to increase machine learning methods' explainability in a broad set of engineering applications.

## Acknowledgements

This research was funded by the Deutsche Forschungsgemeinschaft (DFG, German Research Foundation) – 252408385 – IRTG 2057 and the European Union as part of the European Regional Development Fund and the Commercial Vehicle Cluster Südwest.